\documentclass{ws-ijmpe}

\begin{document}

\markboth{Lacroix, Bender, Duguet}{Configuration mixing within the energy density functional formalism: \\ pathologies and cures.}

\catchline{}{}{}{}{}

\title{Configuration mixing within the energy density functional formalism: \\ pathologies and cures.}

\author{DENIS LACROIX}

\address{GANIL, CEA and IN2P3, Bo\^ite Postale 5027, F-14076 Caen Cedex, France}

\author{MICHAEL BENDER}

\address{Universit{\'e} Bordeaux and CNRS/IN2P3, CENBG, UMR5797,
             F-33175 Gradignan, France}

\author{THOMAS DUGUET}

\address{CEA, Centre de Saclay, IRFU/Service de Physique Nucl\'eaire, F-91191 Gif-sur-Yvette, France}
\maketitle

\begin{history}
\received{(received date)}
\revised{(revised date)}
\end{history}

\begin{abstract}

Configuration mixing calculations performed in terms of the Skyrme/Gogny Energy Density Functional (EDF) rely on
extending the Single-Reference energy functional into non-diagonal EDF kernels.
The standard way to do so, based on an analogy with the pure Hamiltonian case and the use of the generalized Wick theorem, is responsible for the recently observed divergences and steps in Multi-Reference calculations. We summarize here the minimal solution
to this problem recently proposed\cite{Lac08} and applied with success to particle number
restoration\cite{Ben08}. Such a regularization method provides suitable corrections of pathologies for EDF depending
on integer powers of the density. The specific case of fractional powers of the density\cite{Dug08}
is also discussed.
\end{abstract}

\section{Energy Density Functional methods}

The nuclear Energy Density Functional (EDF) method is a unique tool to study static and dynamical properties of nuclei
in a unified framework\cite{bender03b}.  Although the nuclear EDF shares several features with Density Functional Theory\cite{lecturenotesFNM} (DFT), the strategy used is different as it embraces two successive levels of description.

On the first level, traditionally called "self-consistent mean-field theory", Hartree-Fock (HF) or
Hartree-Fock-Bogoliubov (HFB), a single product state $\Phi_0$ provides the normal $\rho^{00}$ and anomalous $\kappa^{00}$ density matrices the many-body energy is a functional of. We call this method a single-reference (SR) EDF approach and denote by
${\cal E}_{SR}[\Phi_0]={\cal E}_{SR}[\rho^{00}, \kappa^{00}, \kappa^{00 \, \ast}]$ the actual EDF. Although such a restriction is not necessary, one usually builds the EDF from an effective vertex (of the Skyrme or Gogny type), whose parameters are adjusted to reproduce a selected set of experimental observations. Independently of the starting point, the EDF can be written as in any arbitrary basis as
\begin{eqnarray}
\label{eq:e00}
{\cal E}_{SR}[\rho^{00}, \kappa^{00}, \kappa^{00 \, \ast}]
& = & \sum_{ij} t_{ij} \, \rho^{00}_{ji}
      + \tfrac{1}{2} \sum_{ijkl} \bar{v}^{\rho\rho}_{ijkl} \,
        \rho^{00}_{ki} \, \rho^{00}_{lj}
      + \tfrac{1}{4} \sum_{ijkl} \bar{v}^{\kappa\kappa}_{ijkl} \,
        \kappa^{00 \, \ast}_{ij} \, \kappa^{00}_{kl} \\
&& +  \tfrac{1}{6} \sum_{ijklmn} \bar{v}^{\rho\rho\rho}_{ijklmn} \,
        \rho^{00}_{li} \, \rho^{00}_{mj} \, \rho^{00}_{nk}
+ \tfrac{1}{4} \sum_{ijklmn} \bar{v}^{\rho\kappa\kappa}_{ijklmn} \, \rho^{00}_{li} \,
        \kappa^{00 \, \ast}_{jk} \, \kappa^{00}_{mn} + \ldots \, \, \, , \nonumber
\end{eqnarray}
where the first term accounts for the uncorrelated kinetic energy, whereas $\bar{v}^{\rho\rho}$, $\bar{v}^{\kappa\kappa}$, $\bar{v}^{\rho\rho\rho}$, \ldots denote effective vertices associated with the different terms of the EDF. There are a few important comments to be made at this point. First, and although it formally resembles it, Eq.~(\ref{eq:e00}) should not be confused with the expectation value of a Hamiltonian containing two-body, three-body, \ldots interactions in the Hartree-Fock-Bogolyubov state $\Phi_0$. For this to be true, specific properties, e.g.\ $\bar{v}^{\rho\rho}_{ijkl}=\bar{v}^{\kappa\kappa}_{ijkl}$ and $\bar{v}^{\rho\rho}_{ijkl}=-\bar{v}^{\rho\rho}_{ijlk}$ for all $(i,j,k,l)$, would have to be satisfied, which is usually not the case in the EDF context. Second, most popular and performing EDF cannot be written under the form of Eq.~(\ref{eq:e00}) as they contain a dependence on a non-integer power of the (local) normal density\cite{bender03b}. We anticipate, however, that future EDFs will be constructed under such a form, typically truncated at forth or fifth power. Indeed, the regularization procedure presented here is inapplicable to EDFs containing non-integer powers of the density matrices\cite{Dug08}.

While static collective correlations, e.g.\ pairing and deformation, can be accounted for within the SR EDF formalism through the symmetry breaking of the auxiliary state $\Phi_0$, dynamical collective correlations requires to perform a so-called Multi-Reference (MR) calculations, traditionally denoted as "beyond-mean-field". Such an extension, built by analogy with the Generator Coordinate Method (GCM) in the Hamiltonian formalism\cite{Rin81}, allows one not only to incorporate additional correlations but also to describe low-energy spectroscopy and transition probabilities between states characterized by symmetry-restored quantum numbers. In strict analogy with the Hamiltonian formalism, the MR EDF is written as
\begin{equation}
\mathcal{E}[\Psi] \equiv \frac{\sum_{\{0,1\} \in {\rm M\!R}} f^{\ast}_{0} \, f_{1} \, \,
        \mathcal{E}_{MR}[\Phi_0,\Phi_1] \,
        \langle \Phi_{0} | \Phi_{1} \rangle}
       {\sum_{\{0,1\} \in {\rm M\!R}} f^{\ast}_{0} \, f_{1} \,
        \langle \Phi_{0} | \Phi_{1} \rangle}
\, \, \, , \label{MRenergy}
\end{equation}
where non-diagonal matrix elements $\langle \Phi_0 | \hat{H} | \Phi_1 \rangle/\langle \Phi_0 | \Phi_1 \rangle$ have been replaced by their
EDF counterpart $\mathcal{E}_{MR}[\Phi_0,\Phi_1]$. The weight functions $f$ are determined either by symmetry considerations, by diagonalization, or both. The product states $\Phi_i$ belonging to the MR set are chosen according to the collective modes one wants to describe. In the absence of a well-founded prescription to build $\mathcal{E}_{MR}[\Phi_0,\Phi_1]$, only specific constraints can be imposed. For a number of reasons\cite{Robledo07a}, it is necessary to impose that $\mathcal{E}_{MR}[\Phi_0,\Phi_0] \equiv \mathcal{E}_{SR}[\Phi_0]$ and $\mathcal{E}_{MR}[\Phi_1,\Phi_0] = \mathcal{E}^*_{MR}[\Phi_0,\Phi_1]$. Following the Hamiltonian formalism, the most natural guidance is provided by the generalized Wick theorem\cite{balian69a} (GWT) which tell us that $\mathcal{E}_{MR}[\Phi_0,\Phi_1]$ is obtained by replacing
SR density matrices by transition ones, i.e. $[\rho^{01}, \kappa^{01}, \kappa^{10 \, \ast}]$, in Eq.~(\ref{eq:e00}). However, we have shown that the use of GWT-based functional energy kernels is the source of the pathologies recently observed in MR-EDF calculations\cite{Lac08,Ben08,Dug08}.

\subsection{Pathologies observed in configuration mixing calculations}

An example of deformation energy surface obtained through a MR calculation based on Particle-Number Restoration (PNR) is given for $^{18}$O in
Fig. \ref{fig:o18:e:pes:c} using the trilinear SIII Skyrme EDF. Starting from a SR-EDF built from the auxiliary state $\Phi_0$ which explicitly breaks the particle-number symmetry, in order to account for static pairing correlations explicitly, dynamical pairing correlations associated with PNR can be incorporated through a MR EDF calculation. Building the MR set from product states rotated in gauge space $| \Phi_\varphi \rangle = e^{i\varphi \hat{N}}| \Phi_0 \rangle$, Eq.~(\ref{MRenergy}) specified to PNR reads\cite{Lac08}
\begin{equation}
\label{scalar2}
\mathcal{E}^{N}
\equiv \int_{0}^{2\pi} \! d\varphi \,
       \frac{e^{-i\varphi N}}
               {2\pi \, c^{2}_{N}} \,
       \mathcal{E}_{MR}[\Phi_0 , \Phi_\varphi] \,  \langle  \Phi_0 | \Phi_{\varphi} \rangle
\, .
\end{equation}
In practice, Eq.~(\ref{scalar2}) is numerically estimated using the Fomenko\cite{fom70a} discretization procedure. The left panel of Fig.~\ref{fig:o18:e:pes:c} presents results obtained for two different numbers of mesh points in the discretization (dotted and dashed lines). Obvious pathologies are visible, i.e.\ (i) the estimate of the energy landscape does not converge and (ii) non-physical steps appear at particular deformations as one increases the number of mesh points used in the Fomenko procedure\footnote{The fact that no divergence occurs is due to the particular form of the functional used which is strictly bilinear in the same isospin\cite{Dug08}.}. Authors have not only also faced the problem
for PNR\cite{Ang01,Ben07} but also when performing angular-momentum restoration\cite{Zdu07}. It has been recognized\cite{Ang01} that divergences in PNR may appear when either a proton or neutron single-particle level crosses the Fermi en energy, as pair of states differing by $\pi/2$ are orthogonal in this case, i.e.\ $\langle  \Phi_0 | \Phi_{\varphi} \rangle = 0$. When the same (density-independent) vertices are used in the p-h and p-p channel and the exchange is properly taken into account, problems were shown to disappear. The problem has been characterized more precisely thanks to a complex plane analysis\cite{Dob07}, demonstrating in particular the less obvious but more profound occurrence of steps. This technique, however, cannot be extended to the restoration of other symmetries and does not lead to a practical solution of the problem.

\section{Minimal solution to the problem}

\begin{figure}[t!]
\centerline{\includegraphics[width=6.0cm]{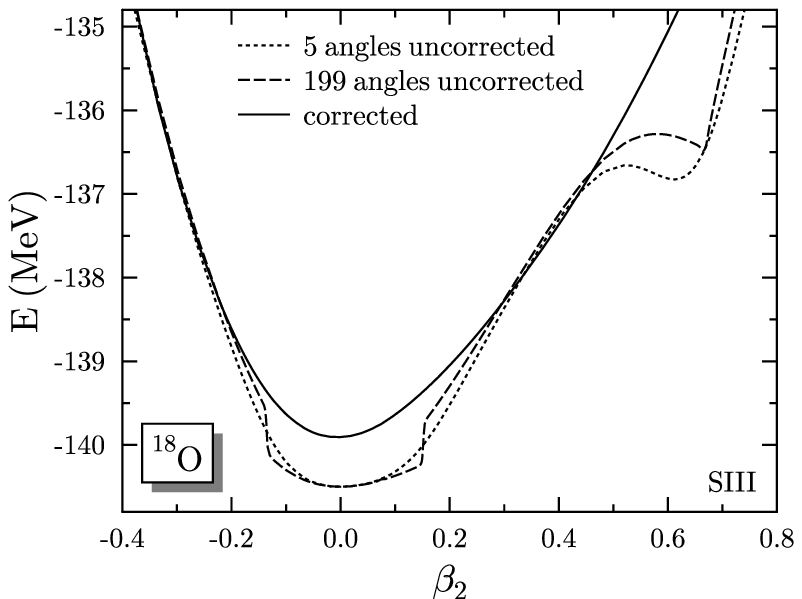}
            \includegraphics[width=6.0cm]{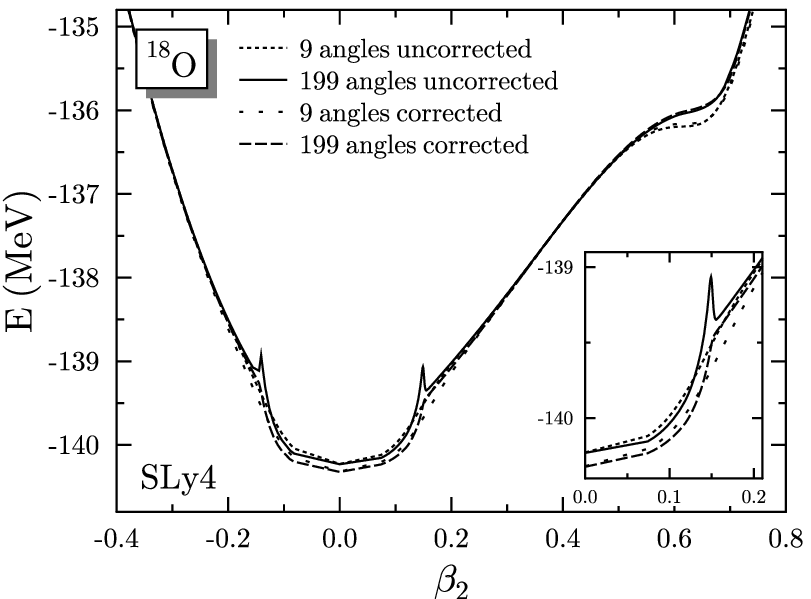}}
\caption{
\label{fig:o18:e:pes:c}
Left: Uncorrected (dotted and dashed lines) and
corrected (solid line) particle-number restored quadrupole deformation energy obtained for $^{18}$O with SIII and calculated
with $L=5$ and 199 discretization points of the integral in gauge space. The two corrected curves are superimposed. Right: attempt to regularize the particle-number restored energy of $^{18}$O obtained with SLy4 that contains a non-integer power of the (local) normal density.
}
\end{figure}

We have recently shown that the origin of difficulties can be traced back to the strategy used to design energy kernels entering the MR-EDF, i.e.\ the use of the GWT as a guidance. An early hypothesis\cite{Ang01} has been confirmed later\cite{Ben07} by avoiding the use of the GWT in the PNR case. 
More recently, a general solution which applies to any type of configuration mixing has been formulated\cite{Lac08}. The technique makes use of the following trick: given a pair of quasi-particle vacua, denoted by
$| \Phi_0 \rangle$ and $| \Phi_1 \rangle$ (with possibly $\langle \Phi_0 | \Phi_1 \rangle=0$), one can always find a  simple "BCS like"
expression connecting these two states, i.e.\cite{Rin81}:
\begin{eqnarray}
\label{eq:BCSqp}
| \Phi_1 \rangle
= \tilde{\cal C}_{01} \,
  \prod_{p > 0}
  \left(   \bar A^*_{pp}
         + \bar B^*_{p\bar p} \,\tilde{\alpha}^+_p \, \tilde{\alpha}^+_{\bar p}
  \right)
  | \Phi_0 \rangle \, \, \,
.
\end{eqnarray}
In the quasi-particle basis where Eq.~(\ref{eq:BCSqp}) is valid, GWT-based energy kernels read, e.g.\ for a strictly bilinear EDF, as (omitting the kinetic term)
\begin{alignat}{3}
{\cal E}_{MR}[\Phi_0,\Phi_1]
=
& %
  \frac{1}{2} \sum_{\nu \mu} \bar{v}^{\rho\rho}_{\varphi_\nu \varphi_\mu \varphi_\nu \varphi_{\mu}} &
&  + \frac{1}{4} \sum_{\nu \mu} \bar{v}^{\kappa\kappa}_{\varphi_\nu \phi_{\bar \nu} \varphi_\mu \phi_{\bar \mu}}
\label{kernel1}
 \\
+ &  \frac{1}{2} \sum_{\nu \mu} \bar{v}^{\rho\rho}_{\varphi_\nu \varphi_\mu \phi_{\nu} \varphi_{\mu}} \,
     \bar Z_{\nu \bar\nu } &
&  + \frac{1}{4} \sum_{\nu \mu} \bar{v}^{\kappa\kappa}_{\varphi_\nu \varphi_{\bar \nu}\varphi_\mu \phi_{\bar \mu}}
\,
     \bar Z_{\nu \bar \nu}
\label{kernel2}
\\
+ &  \frac{1}{2} \sum_{\nu \mu} \bar{v}^{\rho\rho}_{\varphi_\mu \varphi_\nu \varphi_\mu \phi_{ \nu}} \,
     \bar Z_{\nu \bar\nu } &
&  + \frac{1}{4} \sum_{\nu \mu} \bar{v}^{\kappa\kappa}_{\varphi_\mu \phi_{\bar\mu} \phi_{ \nu} \phi_{\bar \nu}} \,
      \bar Z_{\nu \bar \nu}
\label{kernel3}
 \\
+ &  \frac{1}{2} \sum_{\nu \mu} \bar{v}^{\rho\rho}_{\varphi_\nu \varphi_\mu \phi_{\nu} \phi_{\mu}} \,
     \bar Z_{ \nu\bar\nu} \, \bar Z_{ \mu \bar\mu} &
&  + \frac{1}{4} \sum_{\nu \mu} \bar{v}^{\kappa\kappa}_{\varphi_\nu \varphi_{\bar \nu} \phi_{ \mu} \phi_{\bar
\mu}} \,
     \bar Z_{ \nu \bar\nu } \, \bar Z_{ \mu \bar\mu } \, \, , \label{kernel4}
\end{alignat}
where $(\mu,\bar \mu)$ denote a canonical pair in the specific quasi-particle representation,
$\bar Z_{\bar \nu \nu} =(\bar B_{\bar \nu \nu} / \bar A^{-1}_{\nu \nu})^*$ while $\varphi_\nu$ and $\phi_\mu$ stand for the upper and lower components of the quasi-particle states\cite{Lac08}. Using expressions~(\ref{kernel1}-\ref{kernel4}) is convenient as it separates the contributions remaining in the SR limit (line \ref{kernel1}) from the rest. It also allows to identify the different sources of problems.
{\bf (i)} {\bf Self-interaction:} Well-kown from DFT, self-interaction relates to the fact that a particle should not interact with itself,
which, however, happens when the vertices $\bar{v}^{\rho\rho}$ are not anti-symmetrized\cite{Lac08}. This is almost always 
the case in actual EDFs at least because of the
approximations used to treat Coulomb exchange. Such a self-interaction, if present at the SR level, further contributes at the MR level (lines (\ref{kernel2}-\ref{kernel3})). {\bf (ii)} {\bf Self-pairing:} this new concept\cite{Ben08} is specific to EDFs treating pairing correlations explicitly and relates to the fact that two paired particles should not generate correlation energies, beyond their direct interaction, by scattering onto themselves. Again, such a spurious contribution appears at both SR and MR levels.
{\bf (iii)} {\bf Steps and divergences:} As the energy kernel is multiplied by $\langle \Phi_0 | \Phi_1 \rangle \propto \prod_\nu \bar A^{*}_{\nu \nu}$ in the MR energy (see Eq.~(\ref{MRenergy})), only terms with $\nu = \mu$ or $\nu = \bar \mu$ in line~(\ref{kernel4}) can lead to divergences and steps when $\bar A^{*}_{\nu \nu}=0$. In the pure Hamiltonian case, i.e. $\bar{v}^{\rho\rho}_{ijkl}=\bar{v}^{\kappa\kappa}_{ijkl}$ and $\bar{v}^{\rho\rho}_{ijkl}=-\bar{v}^{\rho\rho}_{ijlk}$ for all $(i,j,k,l)$, the dangerous contributions coming from the two terms in Eq.~(\ref{kernel4}) exactly cancel out and no divergence or step occurs. However, when different or non-antisymmetrized vertices are used, as in the EDF context, divergences and/or steps are observed, as on the left panel of Fig.~\ref{fig:o18:e:pes:c}.

The quasi-particle basis introduced above allows one to proceed to the analogy with the Hamiltonian formalism on the basis of the standard Wick theorem rather than on the generalized one. Comparing the results of the two schemes, one proves\cite{Lac08} that terms with $\nu = \mu$ or $\nu = \bar \mu$ in line~(\ref{kernel4}) should be zero in the first place and must be removed altogether. This not only solves problem {\bf (iii)} entirely but also remove finite spurious contributions to the MR energy kernel. Such a regularization technique can be applied to any type of configuration mixing performed in terms of an EDF depending on integer powers of the densities. It has been
successfully applied to PNR\cite{Ben08}, as is exemplified on the left panel of Fig.~\ref{fig:o18:e:pes:c} using the SIII Skyrme EDF. The correction not only removes the dependence on the number of mesh points and the non-physical steps, but also corrects the energy landscape {\it away} from those steps.
The case of EDFs depending on non-integer powers of the density matrix has also been analyzed\cite{Dug08}.
Although divergences can be removed using a variant of the method proposed in Ref.\cite{Lac08}, the complex-plane analysis 
demonstrates that the left-over fractional power $\rho^{\gamma}$ with $0<\gamma<1$
 is ill-defined as it generates cusps in the PNR energy landscape, (see right panel of Fig.~\ref{fig:o18:e:pes:c}). Generally speaking, one cannot use a functional that is multi-valued over the complex plane. This has important consequences on the present and future of EDF methods.

\end{document}